\documentclass{aa}
\usepackage{graphicx}
\usepackage{amsmath}
\usepackage[usenames,dvipsnames]{color}  
\RequirePackage{natbib}
\begin{document}
\title{Plasma properties and Stokes profiles during the lifetime of a photospheric magnetic bright  point}
\titlerunning{Evolution of Magnetic Bright Point properties}
\author{R.L. Hewitt\inst{1}, S. Shelyag\inst{2}, M. Mathioudakis\inst{1}, F.P. Keenan\inst{1}}
\authorrunning{R. Hewitt et al.}
\institute{Astrophysics Research Centre, School of Mathematics and Physics, Queen's University Belfast,
BT7 1NN, Northern Ireland, UK\\
\and
Monash Centre for Astrophysics, School of Mathematical Sciences, Monash University, Clayton, Victoria 3800, Australia}

\date{01.01.01/01.01.01}

\abstract {}
{To investigate the evolution of plasma properties and Stokes parameters in photospheric magnetic bright points using 3D magneto-hydrodynamical simulations and radiative diagnostics of solar granulation.}
{Simulated time-dependent radiation parameters and plasma properties were investigated throughout the evolution of a bright point. Synthetic Stokes profiles for the FeI $630.25~\mathrm{nm}$ line were calculated, which allowed the evolution of the Stokes-$I$ line strength and Stokes-$V$ area and amplitude asymmetries to also be investigated.}
{Our results are consistent with theoretical predictions and published observations describing convective collapse, and confirm this as the bright point formation process. Through degradation of the simulated data to match the spatial resolution of SOT, we show that high spatial resolution is crucial for the detection of changing spectro-polarimetric signatures throughout a magnetic bright point's lifetime. We also show that the signature downflow associated with the convective collapse process is reduced towards zero as the radiation intensity in the bright point peaks, due to the magnetic forces present restricting the flow of material in the flux tube.}
{}

\keywords{Sun: Photosphere -- Sun: Magnetic fields -- Plasmas -- Magnetohydrodynamics (MHD)}

\maketitle

\section{Introduction}

\begin{figure*}
\includegraphics[width=\textwidth]{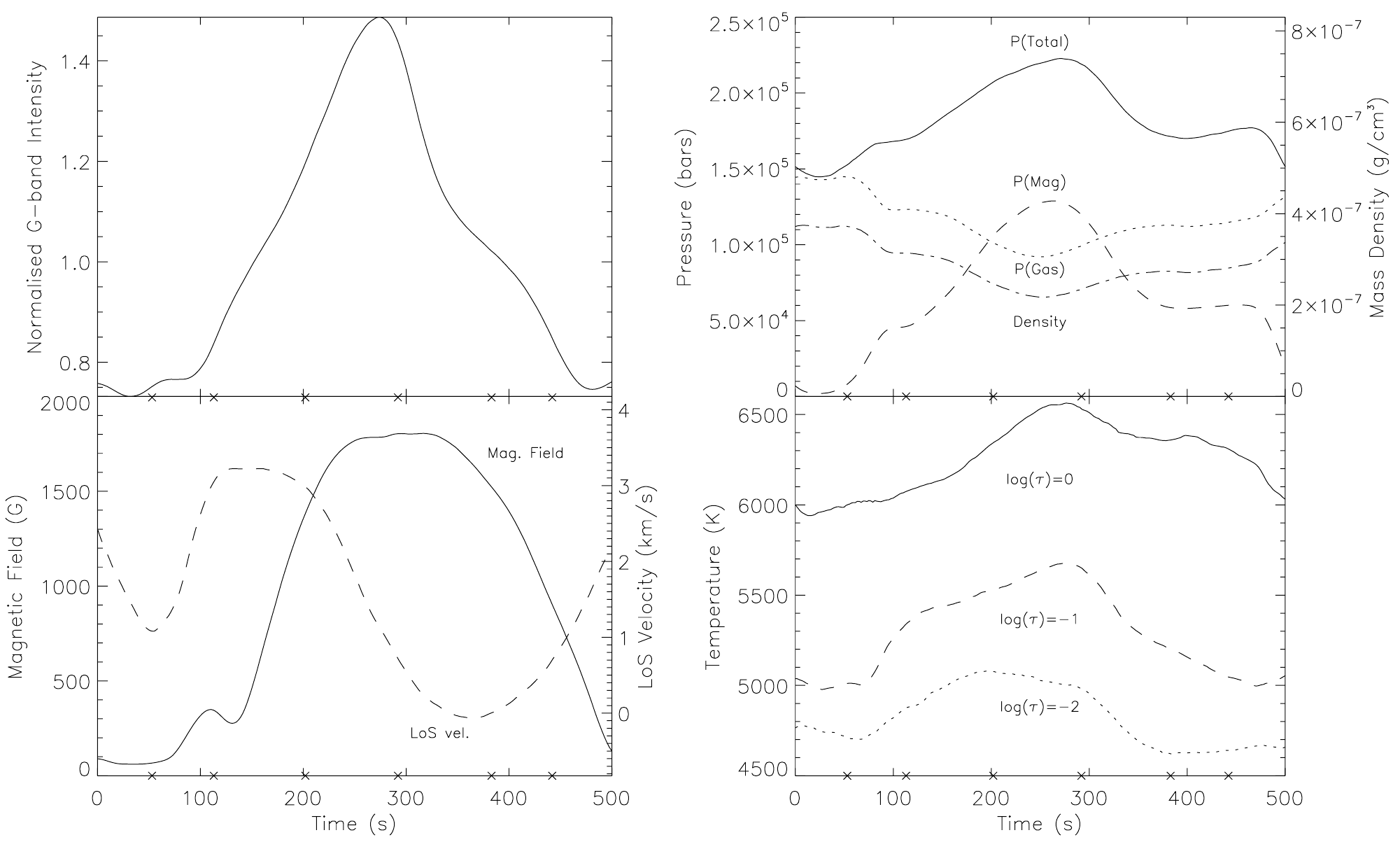}
\caption{Simulated temporal evolution of plasma properties during the formation and disappearance of a MBP. \emph{Top left:} Normalised G-band intensity. \emph{Top right:} Evolution of total pressure (solid), gas pressure (dotted), magnetic pressure (dashed) and mass density (dash-dot) at log($\tau_{500\mathrm{nm}}$)=0. \emph{Bottom left:} Evolution of modulus of the magnetic field (solid) and line-of-sight velocity (dashed) at log($\tau_{500\mathrm{nm}}$)=0, where downward (red-shifted) velocity is positive. \emph{Bottom right:} Plasma temperature evolution at log($\tau_{500\mathrm{nm}}$)=0 (solid),  log($\tau_{500\mathrm{nm}}$)=-1 (dashed) and  log($\tau_{500\mathrm{nm}}$)=-2 (dotted). Markers along the x-axis indicate times at which images and profiles were taken for Figure~\ref{stokes1}.}
\label{t-0}
\end{figure*}

The dominant pattern of the solar photosphere, outside sunspots, is granulation. Granular flows remove magnetic flux from the granules and advect it into the dark intergranular lanes, where the magnetic field accumulates until it reaches a limit ($\sim$500G) given by the equipartition of energy \citep{parker2,bellot1}. However, further intensification occurs by a process called convective collapse. The magnetic field within the intergranular lanes suppresses horizontal convective motions which carry heat to the downflow regions, causing the gas to become cooler and denser, and accelerates the downflow. Due to this, the flux tube becomes partially evacuated. To restore balance between the tube and its surroundings, the tube walls collapse inwards, until stable, and compress the magnetic field lines, thereby intensifying the magnetic field with strengths often in excess of $1~\mathrm{kG}$ \citep{spruit1, takeuchi1, schussler1}. The enhanced magnetic field decreases the density within the flux tube which, in turn, lowers the surface of optical depth unity to layers at higher temperatures. It also leads to increased heating from the granular walls and produces the bright appearance of magnetic bright points (MBPs) \citep{keller1, voegler2, steiner1}.

MBPs evolve on timescales smaller than that of granulation \citep{mehltretter1}, with lifetimes of less than $120~\mathrm{s}$ for almost all MBPs \citep{abramenko1, keys1}. Their diameters range from $\sim120-600~\mathrm{km}$ \citep{bovelet1} with \citet{wiehr1} reporting a predominant diameter of 160$\pm$20 km. Semi-automatic routines for the detection of MBPs in solar observations were recently developed, and statistical studies of their area distribution \citep{crockett1}, transverse velocity distribution \citep{keys1}, were undertaken based on MHD numerical modelling with MuRAM \citep{voegler1} and high-cadence observations with the Rapid Oscillations in Solar Atmosphere (ROSA; \citet{jess2}) instrument at the Dunn Solar Telescope (USA). 
A recent study by \citet{shelyagav} demonstrated the presence of Alfv{\'e}n waves in the intergranular magnetic field concentrations, previously attributed to photospheric vortices \citep{shelyagvort}. These waves carry Poynting flux \citep{shelyagpf}, which potentially is able to propagate and supply the energy to higher layers of the solar atmosphere. While the MBPs exhibit rotational behaviour in the photospheric intergranular lanes, as was observationally demonstrated by \citet{bonetbp}, they do not show ``Alfv{\'e}n-type" transverse oscillatory motions since they are formed deeper than the region where Alfv{\'e}n waves are detected in simulations. 

\citet{berger1} reported on the dynamics of small-scale magnetic elements. However, the first observations which allowed the simultaneous study of plasma properties and magnetic properties simultaneously were undertaken by \citet{nagata1}. They used the Solar Optical Telescope (SOT; \citet{tsuneta1,suematsu1}) onboard Hinode to investigate the evolution of an MBP and confirmed the convective collapse process as the formation mechanism. \citet{narayan1} used observations from the Crisp Imaging SpectroPolarimeter (CRISP; \citet{scharmer2})  installed at the Swedish 1-m Solar Telescope (SST; \citet{scharmer1}), and also made use of spectro-polarimetric data to detect downflows associated with small magnetic features.

While plenty of observations, simulations and analyses of MBPs already exist, there is a need to provide details on the processes of formation and evolution of MBPs in connection to the time-dependent physical and radiative properties of the background solar plasma. In this paper we use radiative MHD simulations of solar magneto-convection to study the temporal evolution of MBPs. Section 2 provides a brief description of the simulations used in this work. Physical parameters such as magnetic field strength, density, line-of-sight velocity, as well as observable quantities such as Stokes parameters are presented in Section 3. A comparison between simulations and published observational findings is also made. Concluding remarks are given in Section 4.

\section{Numerical Simulations}

The simulations employed in this study were produced by the MURaM code \citep{voegler1}. The code solves the 3D radiative MHD equations and simulates the upper convection zone and solar photosphere.
The computational domain has a physical size of $12~$Mm$\times12~$Mm$\times1.4~$Mm which is resolved by 480$\times$480$\times$100 grid cells. This domain has an upper boundary which is treated as closed and is positioned in the upper photosphere, while the lower boundary is located in the upper convection zone and is open to allow plasma to flow freely across the boundary. The visible solar surface, (the layer where most optical radiation is formed), is found to be approximately $600~\mathrm{km}$ below the upper boundary of this domain. Our simulations begin with a non-magnetic model of solar convection until a uniform vertical magnetic field of $200~\mathrm{G}$ is introduced into the domain, and are continued until the models produced are independent of the initial magnetic field and flow configurations. During this time, the magnetic field is expelled from the granules into the intergranular lanes by convective motions, and intergranular magnetic flux concentrations with strengths of up to $2~\mathrm{kG}$ are formed on the solar surface.
An initial field strength of 200G represents a plage region of the solar photosphere and has been found similar to observed plage regions \citep{crockett1,keys1}.

\begin{figure*}
\includegraphics[width=\textwidth]{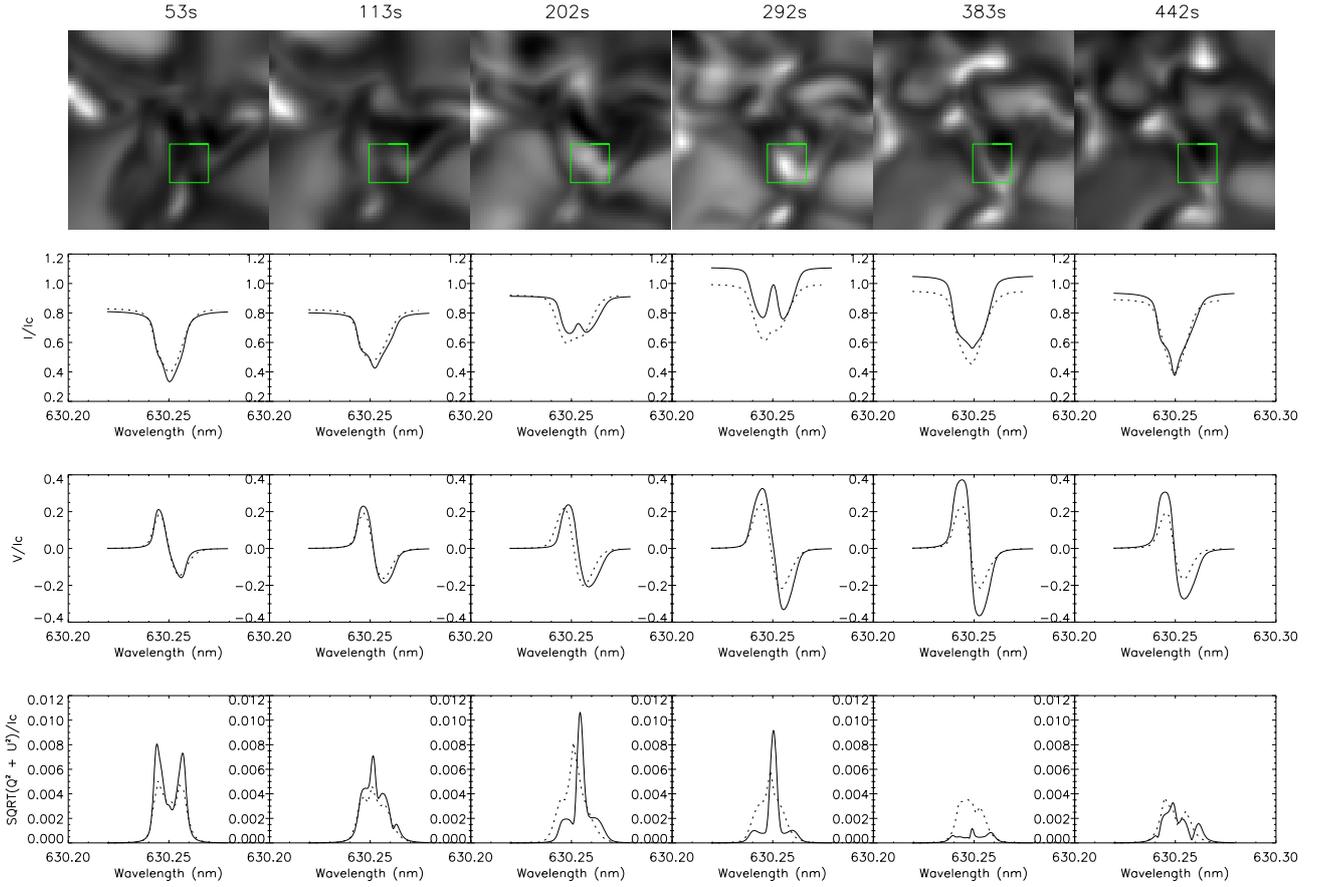}
\caption{\emph{First row:} Sequence of simulated G-band images over time. The green box encloses the area where the investigated MBP is seen to appear and disappear. \emph{Second row:} Sequence of simulated Stokes-$I$ profiles from the centre of the MBP (solid) and these degraded to the SOT spatial resolution (dotted). \emph{Third row:} Sequence of Stokes-$V$ profiles (solid) with these degraded to the SOT spatial resolution (dotted). \emph{Fourth row:} Sequence of Stokes $\sqrt{Q^2+U^2}$ profiles (solid) with these degraded to the SOT spatial resolution (dotted). All profiles are normalised to the average continuum intensity, Ic.}
\label{stokes1}
\end{figure*}

After the intergranular magnetic flux concentrations were formed, we recorded 829 snapshots at a cadence of approximately 2 seconds. This time sequence of the 3D photospheric models was processed by radiative diagnostics routines and stokes parameters for the magnetically-sensitive FeI $630.25~\mathrm{nm}$ absorption line were produced by the STOPRO code \citep{solankiphd,frutigerphd,shelyag1}. Simulated G-band images were also computed, using a simplified routine based on high-resolution opacities provided by \citet{kurucz}, for further analysis.

\section{Temporal evolution of MBPs}

We use numerical simulations of the convection zone and photosphere to investigate the evolution of MBPs. To study the properties of MBPs through their entire lifetime, we selected stable bright points that did not exhibit significant movement along the intergranular lanes during the process of their formation and subsequent disappearance. We examined 829 snapshots, corresponding to almost 30 minutes of real-time simulation, and three MBPs were found suitable for investigation.
However, they all have very similar properties and behaviour and so here we will only discuss in detail one of them. Results from the remaining bright points (referred to as MBP2 and MBP3) are shown in Appendix A and will be referred to in text only to describe differences between the MBPs. Figures~\ref{fig5} and \ref{fig9} show the temporal evolution of plasma properties during the lifetime of a MBP. Figures~\ref{fig6} and \ref{fig10} show the evolution of Stokes profiles throughout the evolution of MBPs with properties of the Stokes-V profiles plotted in figures~\ref{fig7} and \ref{fig11}. Velocity maps for these MBPs are shown in figures~\ref{fig8} and \ref{fig12}. The evolution of the MBP plasma properties are investigated at different heights within the computational domain, namely at log($\tau_{500\mathrm{nm}}$)=0, which will now be referred to as optical depth unity, log($\tau_{500\mathrm{nm}}$)=-1 and log($\tau_{500\mathrm{nm}}$)=-2. 

Figure~\ref{t-0} shows the temporal evolution of G-band intensity, total magnetic field strength, line-of-sight velocity, total pressure, gas pressure, magnetic pressure, density and temperature for one of the MBPs under investigation. The normalised G-band intensity (top-left plot) shows that the MBP has a lifetime of approximately $300~\mathrm{s}$. As it is evident from the comparison of the G-band intensity (top-left plot in Figure~\ref{t-0}) and the magnetic field strength (bottom-left plot), the magnetic field shows good coincidence with the G-band intensity. The peak field strength occurs at approximately the same time as the peak intensity of the MBP. A downflow, reaching velocities in excess of $3~\mathrm{km~s^{-1}}$, appears as the MBP begins to form (dashed curve in bottom-left plot). The peak ($3.2~\mathrm{km~s^{-1}}$) in the downflow velocity occurs before that in the G-band intensity and decreases towards zero as the G-band intensity reaches its maximum.  

\citet{narayan1} obtained spectro-polarimetric data from CRISP at the SST and extracted magnetic field and line-of-sight velocities using Milne-Eddington inversions. Several cases of small-scale transient downflows were detected, ranging from $3.0-5.2~\mathrm{km~s^{-1}}$, associated with small-scale magnetic features which were consistent with observations and simulations describing convective collapse. These downflows, however, were found to temporally coincide with the increase in magnetic field and enhancement in the continuum intensity as opposed to precede them.

\citet{shimizu} used SOT observations and investigated one high-speed downflow ($7.8~\mathrm{km~s^{-1}}$) event that occurred in the quiet Sun. They reported that the high-speed downflow occurred ``almost simultaneously" to a transient brightening in the G-band but preceded the appearance of the magnetic flux concentration.

\citet{nagata1} was the first to observe a convective collapse event. They used SOT observations to report strong downflows reaching $6~\mathrm{km~s^{-1}}$ which occured before the intensity and magnetic field of the MBP peaked. \citet{fischer} also used SOT observations and presented a statistical analysis of 49 convective collapse events. They were all identified by a significant increase in magnetic field preceded by strong photospheric downflows, which ranged from $3.33-3.92~\mathrm{km~s^{-1}}$. These results agree with our simulations, in that the downflow precedes the intensity enhancement and magnetic field strength intensification. This describes the evacuation of the flux tube as the bright point begins to form at the beginning of the convective collapse process.

Figure~\ref{t-0} shows that as the downflow tends to $0~\mathrm{km~s^{-1}}$ just before $300~\mathrm{s}$, the density (dash-dotted curve in the top-right plot) starts to increase again, and the G-band intensity decreases. Once the downflow weakens, the evacuation of the flux tube decreases and the density within the tube increases as material begins to fill the flux tube up again. 

These results, again, agree well with \citet{nagata1} who show the downflow velocity to reach zero while the MBP is still very prominent, and for the intensity to reduce once the downflow ceases. These characteristics are all commonly associated with the convective collapse process \citep{parker1,spruit1}.

The magnetic field exhibits an anti-correlation to the mass density (top-right plot in Figure~\ref{t-0}) over this period, with the lowest density also occurring simultaneously to the G-band intensity peak. As the flux tube fills up with material starting from about $250~\mathrm{s}$, the walls need to expand until they are in equilibrium with their surroundings, decompressing the flux lines and reducing the magnetic field strength.

The evolution of the MBP's temperature at different optical heights in the simulated photosphere was also examined (bottom-right plot in Figure~\ref{t-0}). The temperature dependencies on time appear to follow a similar pattern of increasing and peaking during the G-band intensity maximum before falling back to their pre-bright point level. This is indicative of the changing height of optical depth unity during the MBP evolution. As the flux tube is evacuated, the optical depth unity moves down in the solar atmosphere to layers of higher temperature. Once the downflow reduces, the flux tube density has reached its lowest, and the temperature has reached its highest, the flux tube fills up again moving the optical depth unity higher in the atmosphere and reducing the temperature.

\begin{figure*}
\includegraphics[width=\textwidth]{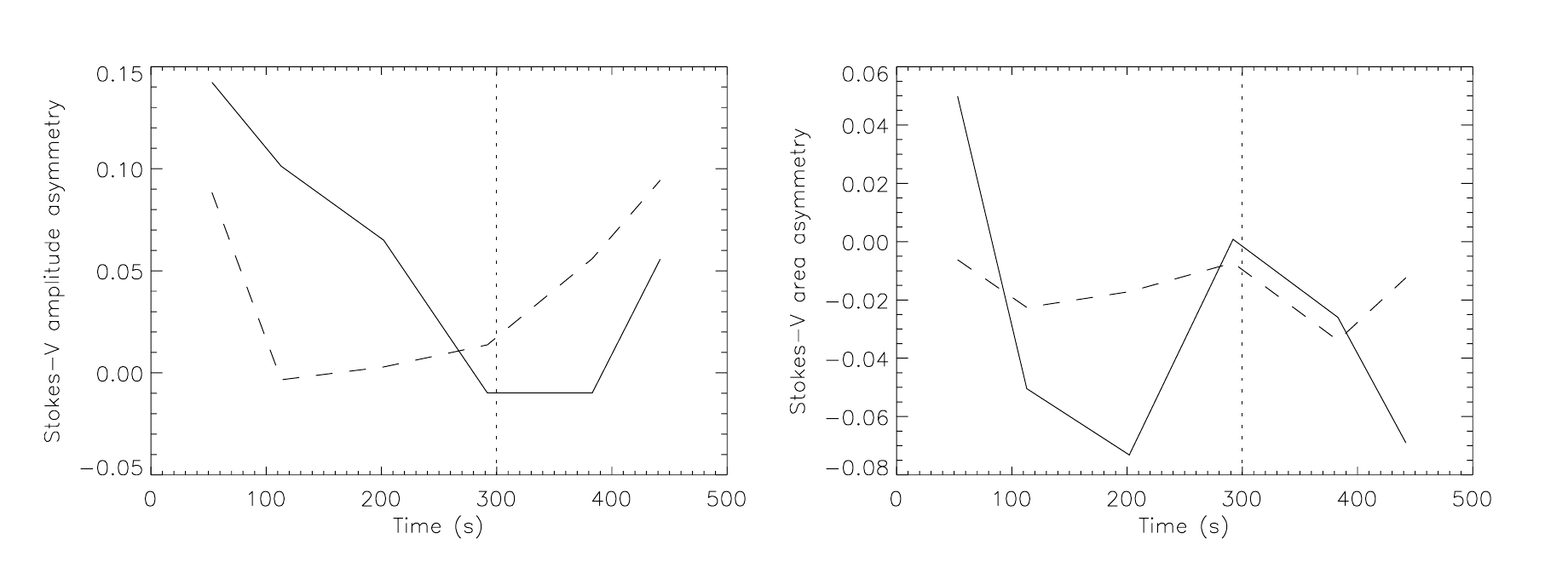}
\caption{\emph{Left:} The Stokes-$V$ amplitude asymmetries emanating from the central pixel of the MBP (solid) for the 6 profiles of Figure~\ref{stokes1}.  \emph{Right:} The corresponding Stokes-$V$ area asymmetries. The asymmetries of the degraded profiles are also shown (dashed).}
\label{asyms}
\end{figure*}

\begin{figure*}
\includegraphics[width=\textwidth]{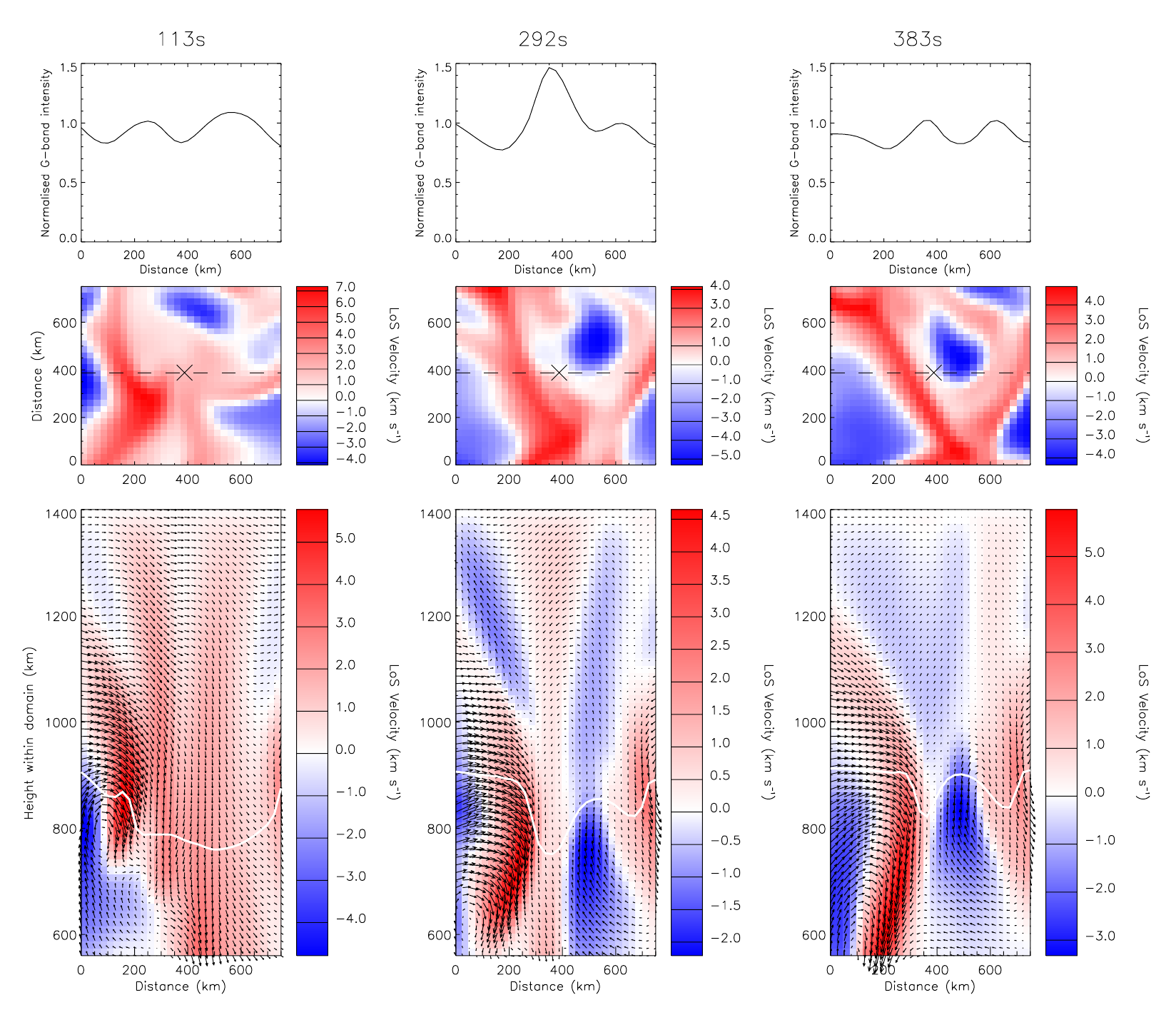}
\caption{\emph{Top row:} Simulations of the normalised G-band intensity slices for three snapshots of the MBP before, during and after its peak intensity. \emph{Middle row:} Line-of-sight velocity maps of the solar surface with the centre of the MBP located at approximately $(375,~375)~\mathrm{km}$. Redshifted velocity is positive. The horizontal dashed line represents the cut at which the G-band intensity plots are taken for the top row graphs. \emph{Bottom row:} Vertical velocity maps perpendicular to the solar surface, taking into account vertical and horizontal velocities with arrows representing the direction and magnitude of velocities. The solid white line is used to show the location of optical depth unity.}
\label{large}
\end{figure*}

Synthesised Stokes profiles of the FeI $630.25~\mathrm{nm}$ line were calculated, as detailed in Section 2, so that the Stokes-$I$ Doppler shift and Stokes-$V$ amplitude and area asymmetries could be analysed. The consistency of these synthetic line profiles when compared with observations is demonstrated by \citet{shelyag1}. Figure~\ref{stokes1} shows the Stokes-$I$, $V$ and $\sqrt{Q^2+U^2}$ profiles at various times during the evolution of the MBP, normalised to the average continuum intensity. Throughout their evolution, the Stokes-$I$ profiles undergo a shift to longer wavelengths as well as Zeeman splitting when the MBP is most prominent. This redshift signifies the strong downflow during the MBP formation and flux tube evacuation. The Zeeman splitting observed in the Stokes-$I$ profiles indicates the strong magnetic field present at the centre of the MBP. However, SOT Stokes-$I$ observations do not show evidence of splitting, despite the MBP being investigated by \citet{nagata1} having a magnetic field strength reaching $2~\mathrm{kG}$. \citet{shimizu} also observed the evolution of Stokes profiles during a magnetic downflow event using SOT however, reported no Zeeman splitting of the Stokes-$I$ profiles. The lack of detectable splitting may be due to the lower spectral and spatial resolution of SOT compared to the simulations. To demonstrate this we degraded the simulated profiles, using an Airy function as an ideal point spread function \citep{wedemeyer}, to match the SOT spatial resolution of 0.3\arcsec and further degraded to match the spectral resolution of 30m\AA (see dotted profiles in Figure~\ref{stokes1}). The degraded profiles show no sign of Zeeman splitting throughout the MBP evolution and are similar to the SOT observations of \citet{nagata1}. However, as can be seen in Appendix A, Figures~\ref{fig6} and ~\ref{fig10}, MBP2 and MBP3 show very weak signs of Zeeman splitting in the degraded Stokes-$I$ profiles. We attribute this difference to the peak magnetic field strengths of $\sim2.3~\mathrm{kG}$ and $\sim2.4~\mathrm{kG}$ respectively, being greater than the other MBP examined. The magnetic field strength distribution of MBPs was found to peak around 1300G \citep{utz}, which would not enable Zeeman splitting detection by SOT. However, as these simulations show, the signature of Zeeman splitting associated with convective collapse may be observable with SOT, if a MBP has a strong enough magnetic field. Therefore, the inconsistency between the simulations and observations appears to be due to the lower resolution of the instrumentation compared to the numerical models.  

The Stokes-$V$ profiles show an increase in amplitude throughout the MBP formation process, which is a result of the line-of-sight magnetic field strength building up during the formation process. The signal becomes stronger when the MBP intensity is at its highest (4th snapshot) but its amplitude continues to grow for a short while even after the MBP's intensity peak (5th snapshot), before decreasing again. The increase in the Stokes-$V$ amplitudes during the formation of the MBP has been previously detected in observations. IBIS spectro-polarimetric data by \citet{viticchie1} found this increase during the coalescence phase of the MBP formation. \citet{shimizu} note a gradual increase in the blue-wing of Stokes-$V$ profiles of magnetic downflow regions, suggesting an increase in magnetic field. \citet{grossmann1} studied magnetic flux concentrations by means of 2D numerical simulations and reported a decrease in the Stokes-$V$ amplitudes after the downflow. Strong upflows were reported to follow strong downflows and so the reduction of Stokes-$V$ amplitudes is suggested to be due to partial cancellation of the Stokes-$V$ lobes due to the presence of opposing velocity flows.

Stokes-$Q$ and Stokes-$U$ are sensitive to the horizontal components of magnetic field, and hence by plotting the square root of the $Q^2+U^2$ profiles we can obtain an indication of the total horizontal magnetic field strength. The last row in Figure~\ref{stokes1} shows $\sqrt{Q^2+U^2}$ profiles during the evolution of the MBP.  The profiles have much weaker signal than the Stokes-V profiles signifying the predominantly vertical orientation of the magnetic field.

The amplitude and area asymmetry of the Stokes-$V$ profiles were also investigated and are shown in Figure~\ref{asyms}. Analysis of the Stokes-$V$ profiles from the simulations revealed significant area and amplitude asymmetries before and after the MBP peak intensity. However, these decrease to zero at the peak of the MBP intensity ($\sim$300s). The Stokes-$V$ asymmetries indicate the presence of magnetic field and velocity gradients \citep{illingasym, solankiasym, shelyag1} during the formation and disappearance of the MBP.

Stokes-$V$ area and amplitude asymmetries were also calculated for the simulated profiles degraded to the SOT resolution (Figure~\ref{asyms}, dotted profiles). These degraded profile asymmetries show less variation throughout the entire lifetime of the MBP. The Stokes-$V$ area asymmetry for degraded simulations remains fairly constant with a low asymmetry of about $-0.02$, while the Stokes-$V$ amplitude asymmetry is also fairly constant with values ranging from about $0.08$ to $0.04$. Therefore, the high spatial resolution available to us through these simulations provides much more detailed information on the Stokes-$V$ asymmetries and therefore of the magnetic field and velocity structures in the solar photosphere.

The velocities related to this MBP are investigated more closely in Figure~\ref{large}. The figure shows velocity maps plotted at three times during the MBP lifetime; before, during and after the MBP G-band intensity peak. These moments are clearly seen in the G-band intensity cuts taken along the MBP. The first and last plots show a fairly constant intensity across the surface. However, the middle plot shows a clearly formed MBP with the intensity peak representing the bright point's centre and the intensity dips on either side indicating the edges of the intergranular lanes in its vicinity. 

The middle row of Figure~\ref{large} shows line-of-sight velocity maps as viewed looking down on the solar surface, with the centre of the MBP at the location $(375,~375)~\mathrm{km}$, while the bottom row are taken from a plane perpendicular to the solar surface that cuts through the centre of the magnetic bright point structure. Arrows plotted over the image represent the direction and magnitude of the horizontal and vertical velocities. The MBP is centred around $375~\mathrm{km}$ across the image with the white line representing the position of optical depth unity. On either side of the MBP, granular flows are seen expelling material from the granules into the intergranular lane with a fairly strong downflow velocity representing the flux tube in the first image.  It is this motion that builds up the magnetic field within the flux tube with the subsequent flux tube evacuation. These are the initial signs of the convective collapse process. As the MBP forms and peaks in the second image, the downflow velocity has reached zero, most notably at the MBP centre just above the optical depth unity level. This can be seen at the MBP centre in the middle row of Figure~\ref{large} and again in Figure~\ref{fig8} and Figure~\ref{fig12}, with stronger downflows surrounding the MBPs. \citet{narayan1} describes downflows that occur at MBP boundaries as opposed to in its centre. The strong kilo-Gauss magnetic field that has built up creates a magnetic pressure that restricts the flow of material in the flux tube and explains the line-of-sight velocity reducing to zero in the MBP centre. In Figure~\ref{large}, the bottom right-hand image shows a small upflow lower in the simulation domain at $\sim(500,~800)~\mathrm{km}$. An upflow of this nature was reported by \citet{nagata1} and was suggested to be a possible ``rebound'', whereby the downflowing material is reflected by the lower layers of the atmosphere and propagates back upwards. These post-convective collapse upflows have also been reported by \citet{bellot1} and \citet{socas1}. In our case, this upflow occurs below optical depth unity and hence was not visible in Figure~\ref{t-0}.

The position of optical depth unity in the bottom row of Figure~\ref{large} (solid white line) changes throughout the lifetime of the MBP. It is lowered at the MBP location when the MBP becomes most prominent. The downflow seen in the bottom left map causes the flux tube to collapse inwards compressing the magnetic field lines and thereby increasing the magnetic field. This magnetic element creates a magnetic pressure and therefore reduced density within the flux tube. The  ``evacuated" flux tube results in an increased photon mean free path and therefore a lower position of optical depth unity. It is this lowering that makes the MBP appear bright when it has formed, as radiation is able to escape from the lower hotter regions.

\section{Concluding Remarks}

In this paper, we investigated the plasma properties and spectro-polarimetric parameters of radiation during the evolution of a photospheric magnetic bright point using numerical simulations and detailed radiative diagnostics of the simulated photospheric magneto-convection models. 
We emphasise that the intergranular lanes are magnetic regions with the MBP being just a feature in a more continuous magnetic field appearing as a result of radiative transfer effects. Our approach in this paper has been to follow, in time,  the area where we see the appearance and disappearance of a MBP and study the flows and magnetic fields in and around that area. We define the MBP feature as an enhancement of the intensity, and magnetic field strength and investigate that location.
Certain aspects of the evolution of the MBP plasma parameters and Stokes profiles in simulation have been analysed such as the magnetic field, line-of-sight velocity, pressure, density and temperature along with the full Stokes vector. 
The evolution of these properties are shown to be consistent with observations of convective collapse \citep{nagata1}.

We observe a strong downflow evacuating the flux tube followed by an enhancement in G-band intensity as  optical depth unity occurs lower in the atmosphere. The density was observed to decrease as a result of the evacuation, causing the flux tube walls to collapse inwards compressing the field lines and intensifying the magnetic field strength . \citet{nagata1}, \citet{fischer} and \citet{shimizu} observed strong downflows associated with MBPs to precede the magnetic field intensification whereas, \citet{narayan1} found the downflows to accompany the magnetic field increase. Our results agree with the former alongside the theory of convective collapse which expects the downflow to precede the magnetic field intensification. We surmise that the results from \citet{narayan1} are probably due to lower temporal resolution.
As the downflow tends towards zero, the evacuation stops thereby starting to increase the density in the tube and causing the magnetic field to begin to reduce as well. This decrease in downflow was noted to be at the MBP centre and caused by strong magnetic pressure that builds up in the flux tube restricting material flow.

The redshift of the Stokes profiles confirm the downflow that occurs during the photospheric MBP formation process, and the Zeeman splitting observed in the Stokes-$I$ profiles indicates the strong magnetic field that builds up throughout its lifetime. However, these signatures of convective collapse are yet to be observed. The increasing amplitude of Stokes-$V$ profiles throughout the process has been observed by \citet{viticchie1} and \citet{shimizu} but the reduction of the amplitude after the process is, as yet, only recorded by \citet{grossmann1} using 2D numerical simulations. The comparison between the Stokes profiles and the plasma properties in Figure~\ref{t-0} allows us to comment on the continued increase in Stokes-$V$ amplitude after the MBP has peaked.

Our results confirm convective collapse as the process involved in the MBP formation. However, the high resolution provided by the simulations allow much more information and structure to be detected and analysed, such as the Zeeman splitting. The spatial resolution of current instrumentation also misses some asymmetry structure in the Stokes-$V$ profiles which provides important information on the line-of-sight velocity and magnetic field gradients. The velocity maps show no downflow in the MBP centre as it peaks, which has not yet been observed. Therefore, high spatial and temporal resolution observations combined with vector magnetograms are essential for understanding the small-scale photospheric structures and dynamics. The Advanced Technology Solar Telescope (ATST; \citet{rimmele}) will detect and track myriads of individual flux concentrations down to a size of 25 km and provide insights on the energy hidden in the dark internetwork field of the quiescent Sun.

\section*{Acknowledgements}
This work is supported by the UK Science and Technology Facilities Council. RLH would like to thank the Department of Education and Learning of Northern Ireland for the award of a PhD studentship. This research was undertaken with the assistance of resources provided at the NCI National Facility systems at the Australian National University, supported by Astronomy Australia Limited, and at the Multi-modal Australian ScienceS Imaging and Visualisation Environment (MASSIVE) (www.massive.org.au).  The authors also thank the Centre for Astrophysics \& Supercomputing of Swinburne University of Technology (Australia) for the computational resources provided. Dr Shelyag is the recipient of Australian Research Council's Future Fellowship (project number FT120100057).

\clearpage
\onecolumn
\begin{appendix}
\section{Results from MBP2 and MBP3}

\begin{figure}[h]
\includegraphics[width=\textwidth]{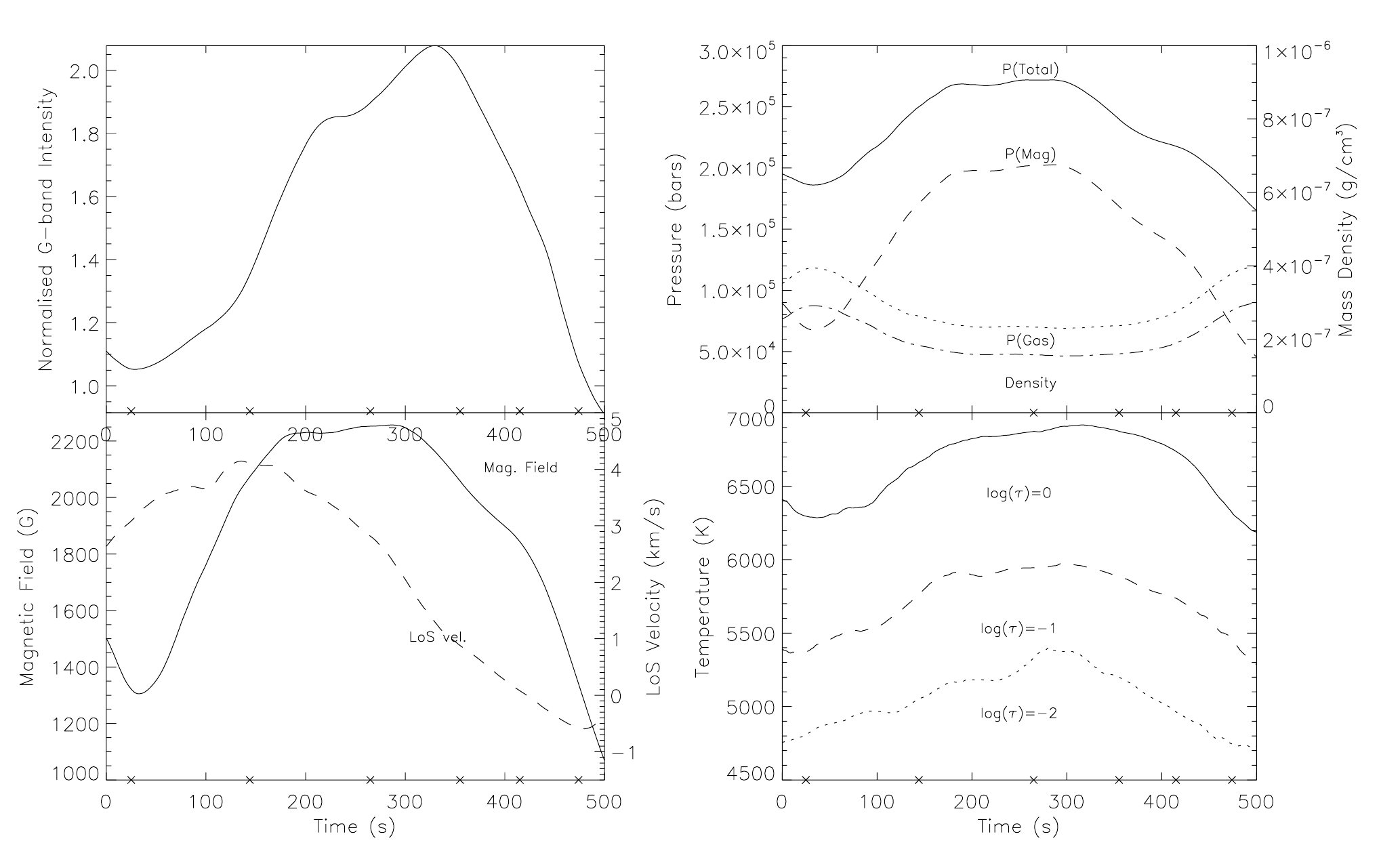}
\caption{Simulated temporal evolution of plasma properties during the formation and disappearance of  MBP2. \emph{Top left:} Normalised G-band intensity. \emph{Top right:} Evolution of total pressure (solid), gas pressure (dotted), magnetic pressure (dashed) and mass density (dash-dot) at log($\tau_{500\mathrm{nm}}$)=0. \emph{Bottom left:} Evolution of modulus of the magnetic field (solid) and line-of-sight velocity (dashed) at log($\tau_{500\mathrm{nm}}$)=0, where downward (red-shifted) velocity is positive. \emph{Bottom right:} Plasma temperature evolution at log($\tau_{500\mathrm{nm}}$)=0 (solid),  log($\tau_{500\mathrm{nm}}$)=-1 (dashed) and  log($\tau_{500\mathrm{nm}}$)=-2 (dotted). Markers along the x-axis indicate times at which images and profiles were taken for Figure~\ref{fig6}.}
\label{fig5}
\end{figure}

\begin{figure*}
\includegraphics[width=\textwidth]{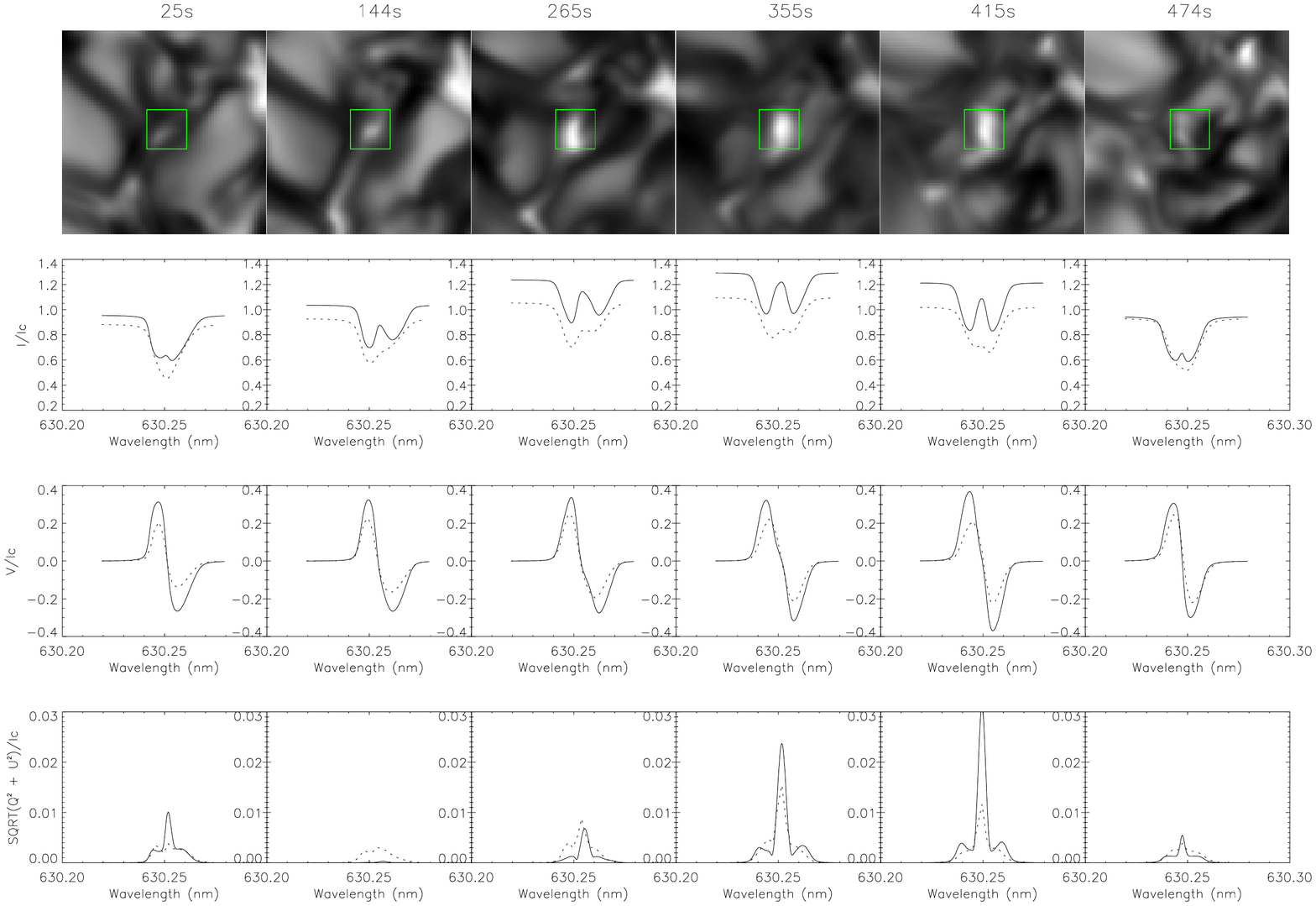}
\caption{\emph{First row:} Sequence of simulated G-band images over time. The green box encloses the area where MBP2 is seen to appear and disappear. \emph{Second row:} Sequence of simulated Stokes-$I$ profiles from the centre of MBP2 (solid) and these degraded to the SOT spatial resolution (dotted). \emph{Third row:} Sequence of Stokes-$V$ profiles (solid) with these degraded to the SOT spatial resolution (dotted). \emph{Fourth row:} Sequence of Stokes $\sqrt{Q^2+U^2}$ profiles (solid) with these degraded to the SOT spatial resolution (dotted). All profiles are normalised to the average continuum intensity, Ic.}
\label{fig6}
\end{figure*}

\begin{figure*}
\includegraphics[width=\textwidth]{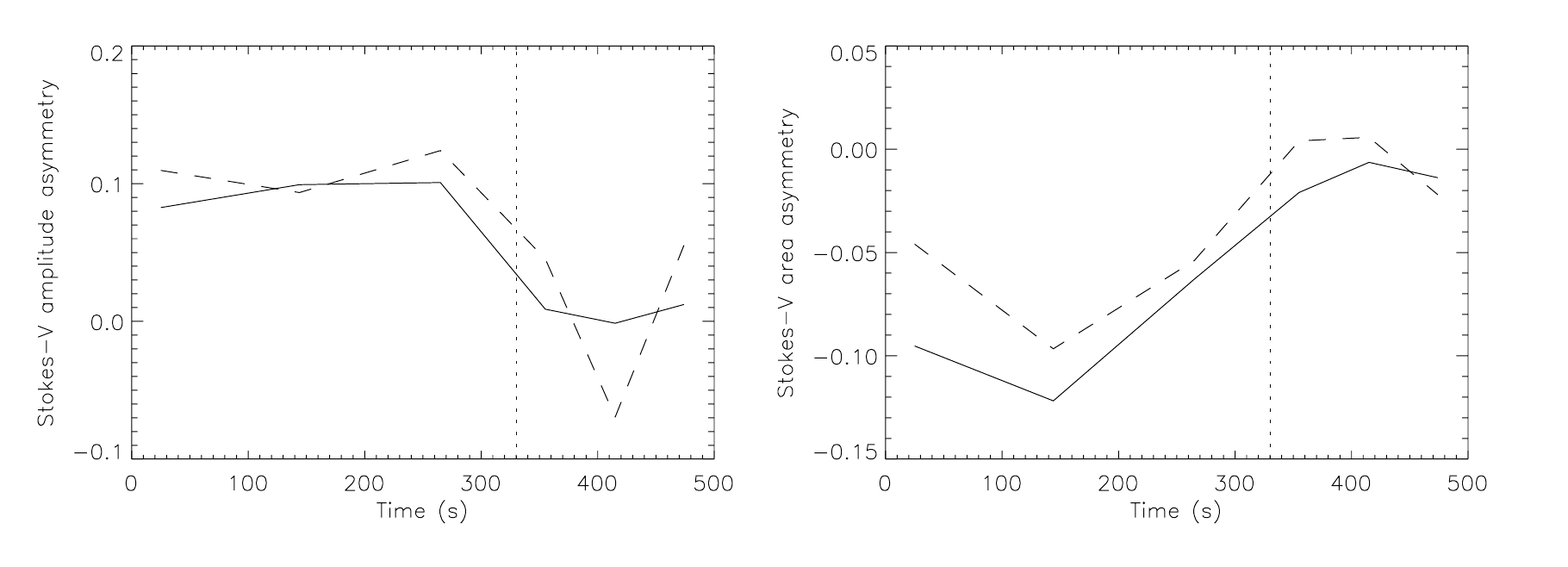}
\caption{\emph{Left:} The Stokes-$V$ amplitude asymmetries emanating from the central pixel of MBP2 (solid) for the 6 profiles of Figure~\ref{fig6}.  \emph{Right:} The corresponding Stokes-$V$ area asymmetries. The asymmetries of the degraded profiles are also shown (dashed).}
\label{fig7}
\end{figure*}

\begin{figure*}
\includegraphics[width=\textwidth]{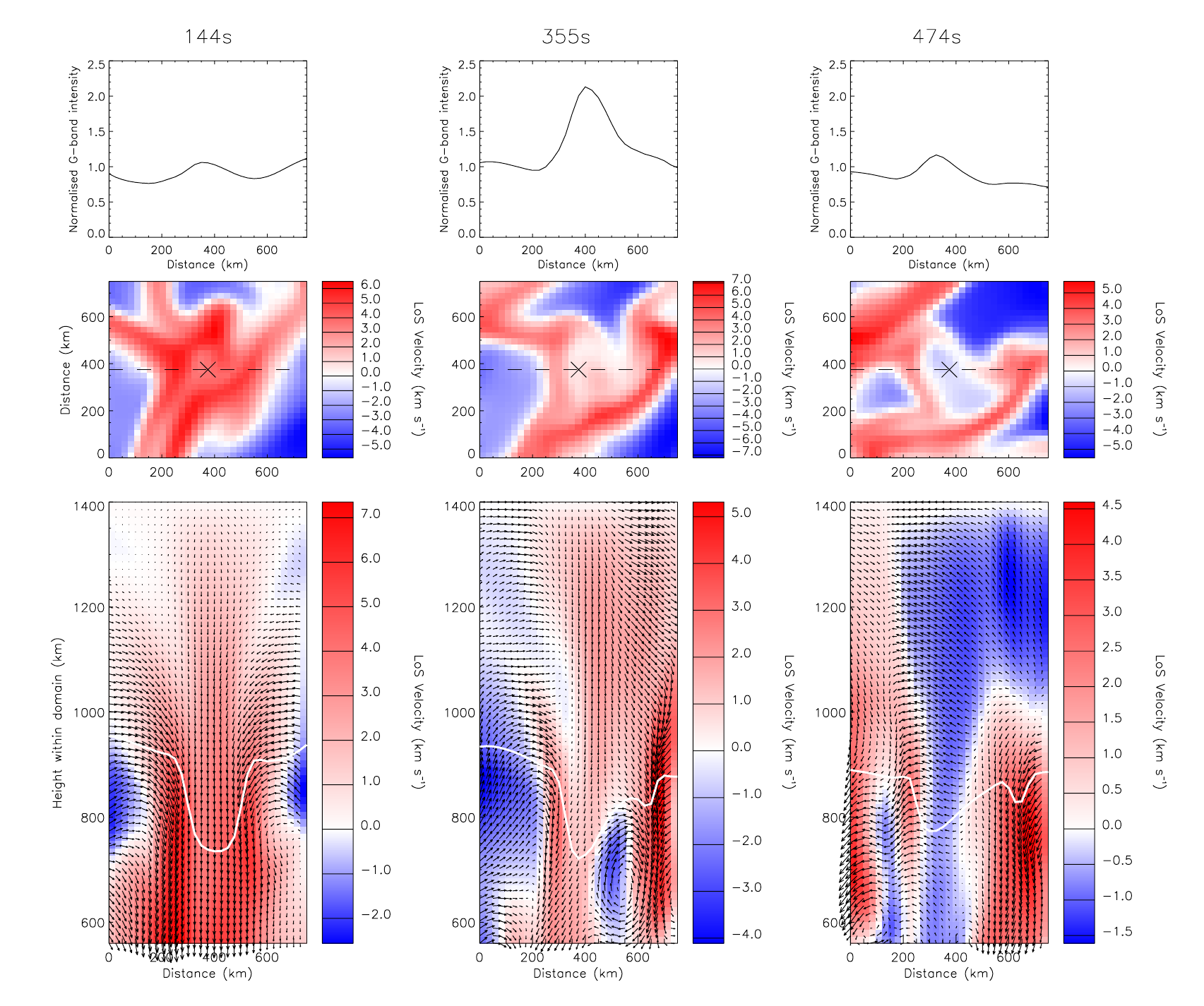}
\caption{\emph{Top row:} Simulations of the normalised G-band intensity slices for three snapshots of the MBP before, during and after its peak intensity. \emph{Middle row:} Line-of-sight velocity maps of the solar surface with the centre of the MBP located at approximately $(375,~375)~\mathrm{km}$. Redshifted velocity is positive. The horizontal dashed line represents the cut at which the G-band intensity plots are taken for the top row graphs. \emph{Bottom row:} Vertical velocity maps perpendicular to the solar surface, taking into account vertical and horizontal velocities with arrows representing the direction and magnitude of velocities. The solid white line is used to show the location of optical depth unity.}
\label{fig8}
\end{figure*}

\begin{figure*}
\includegraphics[width=\textwidth]{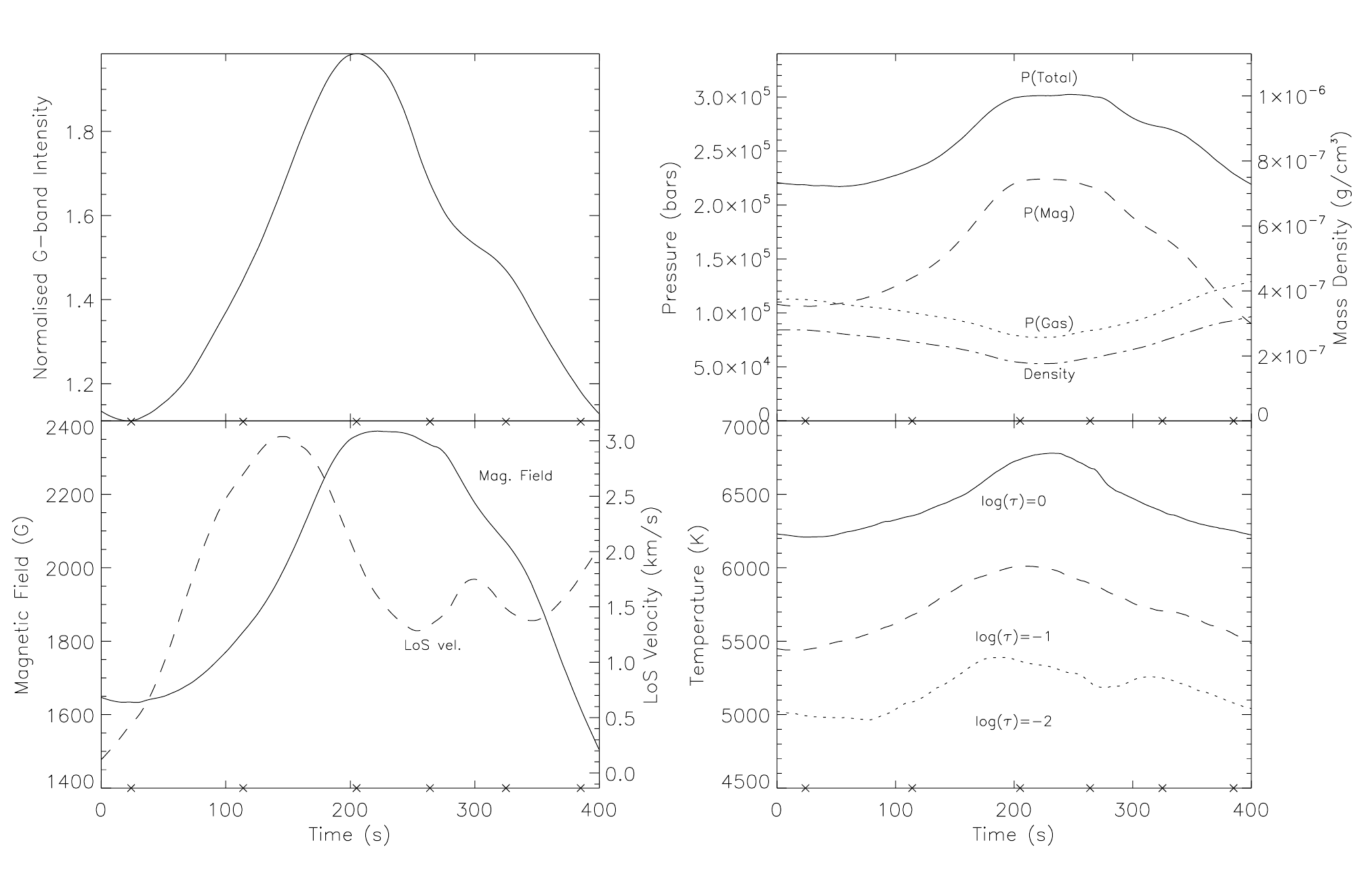}
\caption{Simulated temporal evolution of plasma properties during the formation and disappearance of MBP3. \emph{Top left:} Normalised G-band intensity. \emph{Top right:} Evolution of total pressure (solid), gas pressure (dotted), magnetic pressure (dashed) and mass density (dash-dot) at log($\tau_{500\mathrm{nm}}$)=0. \emph{Bottom left:} Evolution of modulus of the magnetic field (solid) and line-of-sight velocity (dashed) at log($\tau_{500\mathrm{nm}}$)=0, where downward (red-shifted) velocity is positive. \emph{Bottom right:} Plasma temperature evolution at log($\tau_{500\mathrm{nm}}$)=0 (solid),  log($\tau_{500\mathrm{nm}}$)=-1 (dashed) and  log($\tau_{500\mathrm{nm}}$)=-2 (dotted). Markers along the x-axis indicate times at which images and profiles were taken for Figure~\ref{fig10}.}
\label{fig9}
\end{figure*}

\begin{figure*}
\includegraphics[width=\textwidth]{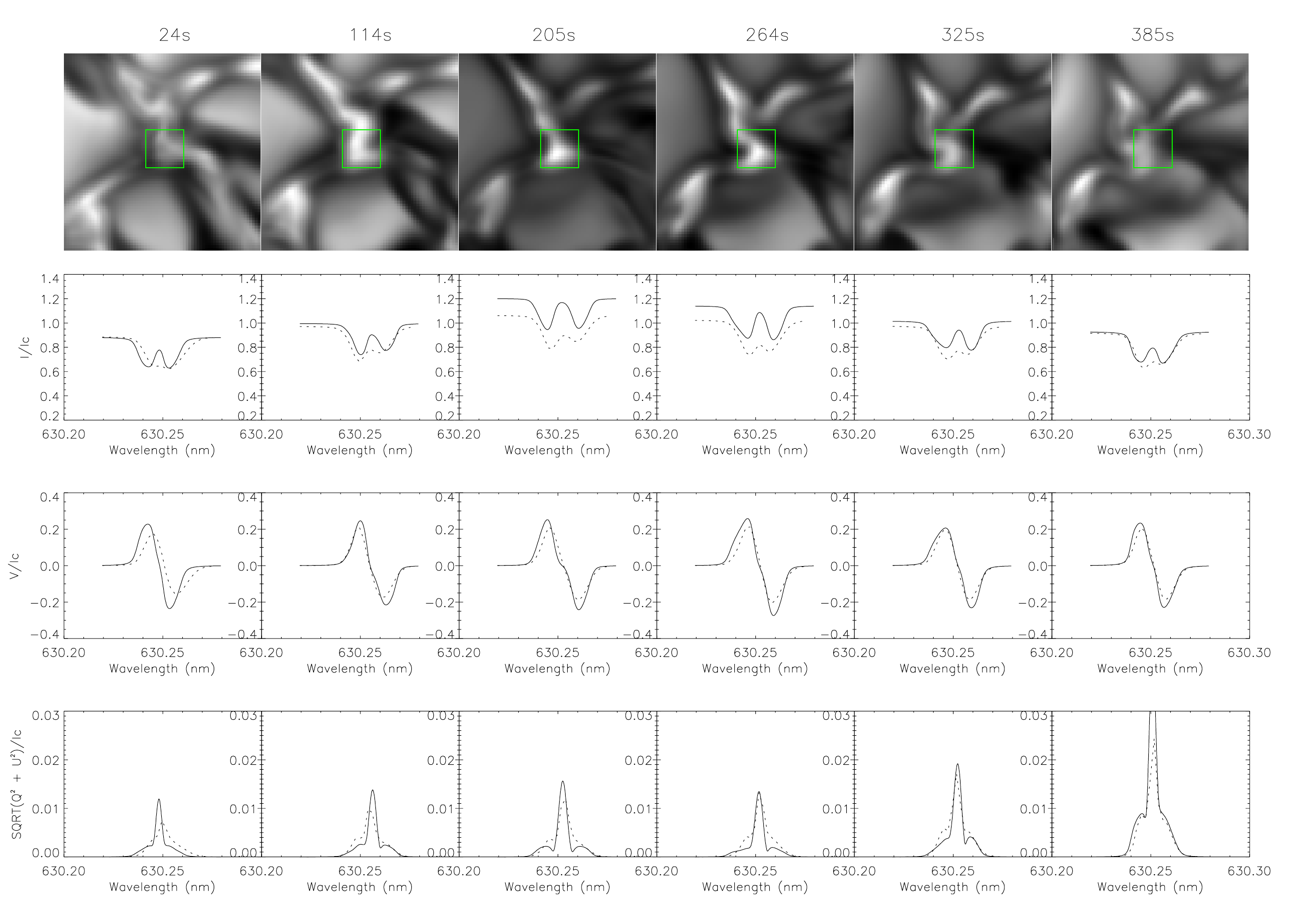}
\caption{\emph{First row:} Sequence of simulated G-band images over time. The green box encloses the area where MBP3 is seen to appear and disappear. \emph{Second row:} Sequence of simulated Stokes-$I$ profiles from the centre of MBP3 (solid) and these degraded to the SOT spatial resolution (dotted). \emph{Third row:} Sequence of Stokes-$V$ profiles (solid) with these degraded to the SOT spatial resolution (dotted). \emph{Fourth row:} Sequence of Stokes $\sqrt{Q^2+U^2}$ profiles (solid) with these degraded to the SOT spatial resolution (dotted). All profiles are normalised to the average continuum intensity, Ic.}
\label{fig10}
\end{figure*}

\begin{figure*}
\includegraphics[width=\textwidth]{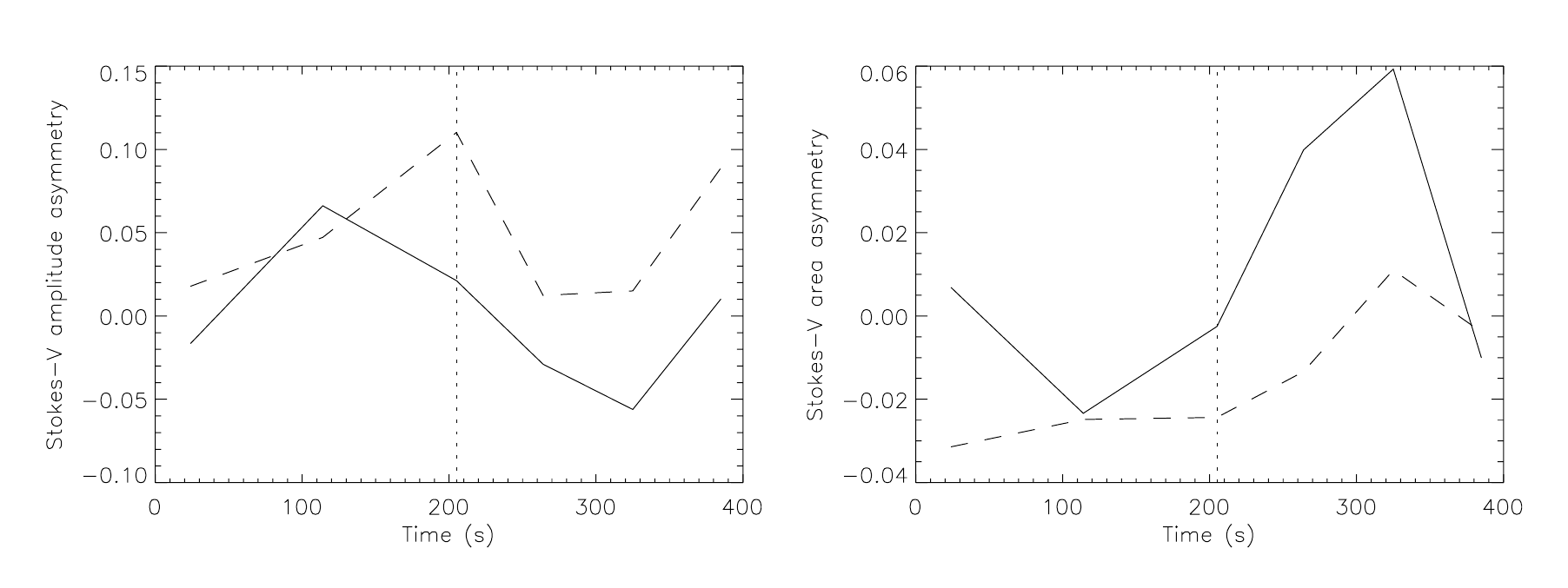}
\caption{\emph{Left:} The Stokes-$V$ amplitude asymmetries emanating from the central pixel of MBP3 (solid) for the 6 profiles of Figure~\ref{fig10}.  \emph{Right:} The corresponding Stokes-$V$ area asymmetries. The asymmetries of the degraded profiles are also shown (dashed).}
\label{fig11}
\end{figure*}

\begin{figure*}
\includegraphics[width=\textwidth]{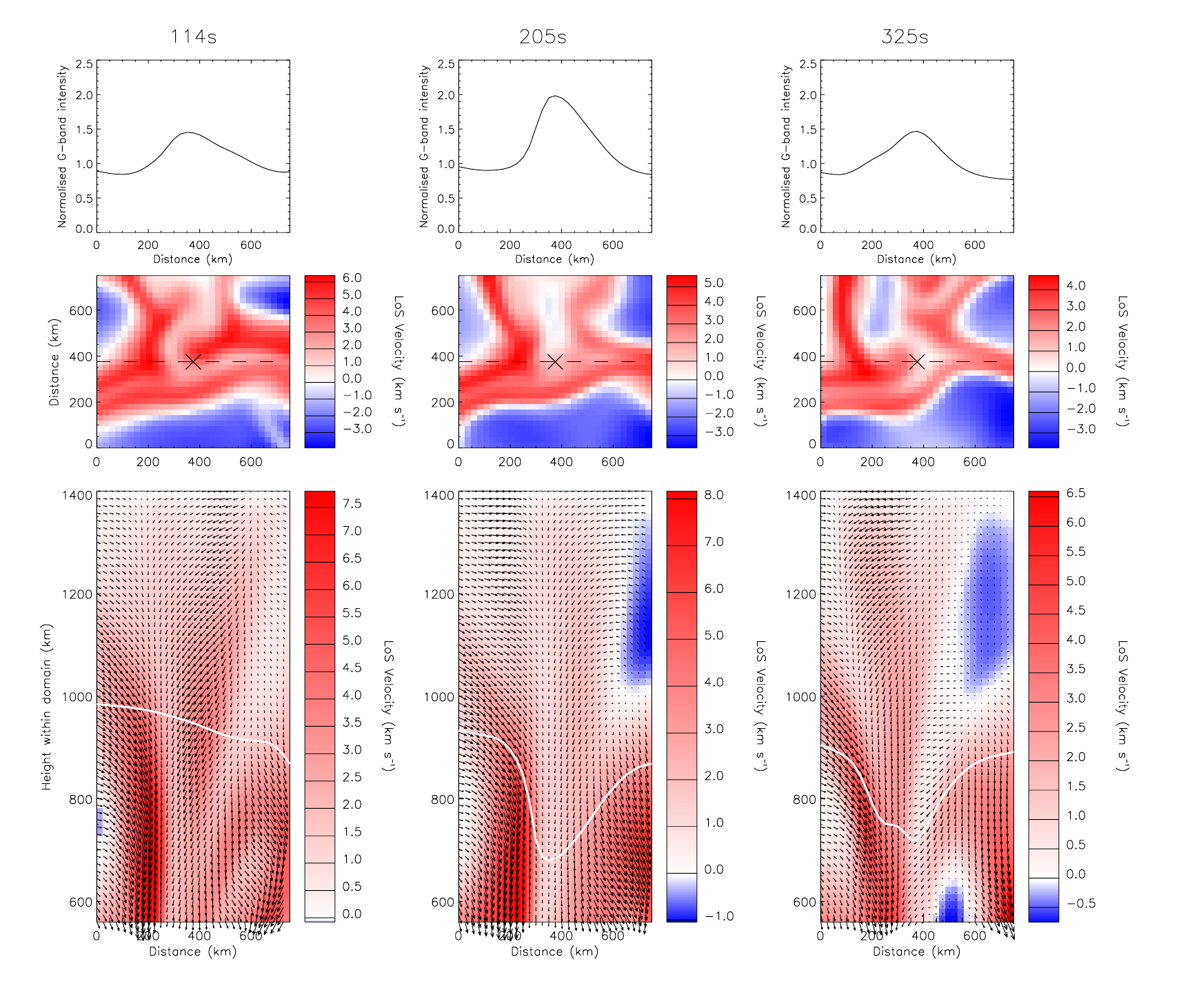}
\caption{\emph{Top row:} Simulations of the normalised G-band intensity slices for three snapshots of the MBP before, during and after its peak intensity. \emph{Middle row:} Line-of-sight velocity maps of the solar surface with the centre of the MBP located at approximately $(375,~375)~\mathrm{km}$. Redshifted velocity is positive. The horizontal dashed line represents the cut at which the G-band intensity plots are taken for the top row graphs. \emph{Bottom row:} Vertical velocity maps perpendicular to the solar surface, taking into account vertical and horizontal velocities with arrows representing the direction and magnitude of velocities. The solid white line is used to show the location of optical depth unity.}
\label{fig12}
\end{figure*}

\end{appendix}

\end{document}